\documentclass[journal,twocolumn,10pt]{IEEEtran}
\usepackage{graphicx,times,amsmath,amssymb,cite,subfigure,url}
\usepackage[lined,ruled,commentsnumbered]{algorithm2e}
\IEEEoverridecommandlockouts \allowdisplaybreaks 

\begin{document}

\title{EEG-Based User Reaction Time Estimation\\ Using Riemannian Geometry Features}

\author{\IEEEauthorblockN{Dongrui Wu\IEEEauthorrefmark{1}, \textit{Senior Member, IEEE}, Brent J. Lance\IEEEauthorrefmark{2}, \textit{Senior Member, IEEE}, \\ Vernon J. Lawhern\IEEEauthorrefmark{2}\IEEEauthorrefmark{3}, \textit{Member, IEEE}, Stephen Gordon\IEEEauthorrefmark{4},\\ Tzyy-Ping Jung\IEEEauthorrefmark{5}, \textit{Fellow, IEEE}, Chin-Teng Lin\IEEEauthorrefmark{6}, \textit{Fellow, IEEE}} \\
\IEEEauthorblockA{\IEEEauthorrefmark{1}DataNova, NY USA}\\
\IEEEauthorblockA{\IEEEauthorrefmark{2}Human Research and Engineering Directorate, U.S. Army Research Laboratory, Aberdeen Proving Ground, MD USA}\\
\IEEEauthorblockA{\IEEEauthorrefmark{3}Department of Computer Science, University of Texas at San Antonio, San Antonio, TX USA}\\
\IEEEauthorblockA{\IEEEauthorrefmark{4}DCS Corp, Alexandria, VA USA}\\
\IEEEauthorblockA{\IEEEauthorrefmark{5}Swartz Center for Computational Neuroscience \& Center for Advanced Neurological Engineering,  University of California San Diego, La Jolla, CA}\\
\IEEEauthorblockA{\IEEEauthorrefmark{6}Centre for Artificial Intelligence, Faculty of Engineering and Information Technology, University of Technology Sydney, Australia}\\
E-mail: drwu09@gmail.com, brent.j.lance.civ@mail.mil, vernon.j.lawhern.civ@mail.mil, sgordon@dcscorp.com, jung@sccn.ucsd.edu, Chin-Teng.Lin@uts.edu.au}
\maketitle

\begin{abstract}
Riemannian geometry has been successfully used in many brain-computer interface (BCI) classification problems and demonstrated superior performance. In this paper, for the first time, it is applied to BCI regression problems, an important category of BCI applications. More specifically, we propose a new feature extraction approach for Electroencephalogram (EEG) based BCI regression problems: a spatial filter is first used to increase the signal quality of the EEG trials and also to reduce the dimensionality of the covariance matrices, and then Riemannian tangent space features are extracted. We validate the performance of the proposed approach in reaction time estimation from EEG signals measured in a large-scale sustained-attention psychomotor vigilance task, and show that compared with the traditional powerband features, the tangent space features can reduce the root mean square estimation error by 4.30-8.30\%, and increase the estimation correlation coefficient by 6.59-11.13\%.
\end{abstract}


\section{Introduction}

Brain-computer interfaces (BCIs) can use brain signals such as the scalp electroencephalogram (EEG) to enable people to communicate or control external devices \cite{McFarland2010,He2015}. Thus, they can help people with devastating neuromuscular disorders such as amyotrophic lateral sclerosis, brainstem stroke, cerebral palsy, and spinal cord injury \cite{Wolpaw2006}. However, there are still many challenges in their transition from laboratory settings to real-life applications, including the reliability and convenience of the sensing hardware \cite{Liao2012}, and the availability of high-performance and robust algorithms for signal analysis and interpretation \cite{Lotte2015,Makeig2012,Wang2015,Jayaram2016}. This paper focuses on the latter, particularly, feature extraction for EEG-based BCIs.

Riemannian geometry (RG) \cite{Lee2002,Berger2007,Pennec2006,Amari2000,Tuzel2008} is a very useful mathematical tool in machine learning and signal/image processing, due to its utility in generating smooth manifolds from intrinsically nonlinear data spaces. Recently it has also been introduced into the BCI community and demonstrated superior performance in a number of applications \cite{Barachant2014,Lotte2015,Barachant2014b,Congedo2013,Barachant2012,Li2012a,Barachant2013,Kalunga2016,Navarro-Sune2016,Yger2015,Waytowich2016}.

For example, Li, Wong, and de Bruin \cite{Li2012a} used RG of the EEG power spectral density matrices for sleep pattern classification. They also proposed a closed-form weighting matrix for the power spectral density matrices to minimize the distance between similar features and to maximize the distance between dissimilar features, and demonstrated better performance than the Euclidian distance and the Kullback-Leibler distance. Barachant et al. \cite{Barachant2012} proposed two RG approaches for motor imagery classification. The first uses the spatial covariance matrices of the EEG signal as features and RG to directly classify them in the manifold of symmetric and positive definite (SPD) matrices. The second maps the covariance matrices onto the Riemannian tangent space, which is a Euclidean space, and then performs variable selection and classification. They achieved comparable or better performance than a multiclass Common Spatial Pattern (CSP) plus Linear Discriminant Analysis (LDA) approach. In \cite{Congedo2013}, Congedo, Barachant, and Andreev further used RG to build calibrationless BCI systems for applications based on event-related potentials, sensorimotor (mu) rhythms, and steady-state evoked potential. It outperformed several state-of-the-art approaches, including xDAWN, stepwise LDA, CSP+LDA, and blind source separation plus logistic regression. Barachant \cite{Barachant2014b} also proposed a spatial filter to increase the signal to signal-plus-noise ratio of magnetoencephalography (MEG) signals before constructing a special form of a covariance matrix for RG feature extraction, and a $k$-means clustering like unsupervised learning algorithm in the Riemannian manifold to improve the offline classification performance. This approach outperformed 266 other approaches and won the Kaggle ``DecMeg2014 -- Decoding the Human Brain" competition\footnote{https://www.kaggle.com/c/decoding-the-human-brain.}, which aimed to predict visual stimuli from MEG recordings of human brain activity. Kalunga et al. \cite{Kalunga2016} proposed an online classification approach in the Riemannian space and showed that it outperformed Canonical Correlation Analysis in Steady-State Visually Evoked Potential classification. Yger, Lotte, and Sugiyama \cite{Yger2015} empirically compared several covariance matrix averaging methods for EEG signal classification. They showed that RG for averaging covariance matrices improved performances for small dimensional problems, but as the dimensionality of the covariance matrix increased, RG became less efficient. Lotte \cite{Lotte2015} also proposed a framework to combine transfer learning, ensemble learning, and RG for calibration time reduction, which outperformed CSP+LDA. The Riemannian distance was used in regularization to emphasize auxiliary users whose covariance matrices are close to the target user. Navarro-Sune et al. \cite{Navarro-Sune2016} proposed a BCI to automatically detect patient-ventilator disharmony from EEG signals. RG of EEG covariance matrices was used in semi-supervised learning for effective classification of respiratory state, and it outperformed the Euclidean distance. Waytowich et al. \cite{Waytowich2016} proposed an approach to integrate RG with transfer learning and spectral meta-learner \cite{Parisi2014}, an offline ensemble fusion approach, for user-independent BCI, and demonstrated in single-trial event-related potential classification that it can significantly outperform existing calibration-free techniques and traditional within-subject calibration techniques when limited data is available.

All above approaches focused on EEG classification problems in BCI, whereas BCI regression problems have been largely overlooked. In theory a regression problem is equivalent to a classification problem with infinitely many classes, and hence the output has much finer granularity than a traditional two-class or multi-class classification problem, which provides richer information in decision making. There are at least two types of BCI regression problems in the literature and practice. The first type is behavioral or cognitive status prediction, e.g., estimating the continuous value of a driver's drowsiness from the EEG \cite{drwuaBCI2015,Lin2008,Lin2005d,drwuTFS2016,drwuEBMAL2016,Lin2006,Wei2015,drwuSMLR2016}, and estimating a subject's response speed in a psychomotor vigilance task (PVT) from the EEG \cite{drwuSF2017}. The second type is direct control applications, e.g., controlling the movement of a mouse cursor using BCI \cite{Wolpaw1991,Wolpaw2000,McFarland1997a,Fruitet2010,Bradberry2011}, and controlling the continuous movement of a hand in the 3D space using EEG \cite{Bradberry2010}.

Once the EEG signal is acquired, the regression problem involves three steps: 1) signal processing to increase the signal-to-noise ratio. Frequency domain filters, such as band pass filters and notch filters \cite{Bradberry2010,Bradberry2011}, and spatial filters, such as independent component analysis \cite{Lin2005d} and CSP \cite{drwuSF2017}, are frequently used here. 2) feature extraction to construct meaningful predictors, e.g., standardized difference of the EEG voltage \cite{Bradberry2010,Bradberry2011}, and EEG power band features \cite{drwuEBMAL2016,drwuaBCI2015,drwuSF2017,drwuSMLR2016}. 3) regression algorithms to estimate the continuous output, e.g., ordinary linear regression  \cite{Bradberry2010,Bradberry2011}, ridge regression \cite{drwuaBCI2015}, LASSO \cite{drwuSF2017}, $k$-nearest neighbors (kNN) \cite{drwuSF2017}, fuzzy neural networks \cite{Lin2006}, transfer learning \cite{drwuTFS2016,Wei2015}, active learning \cite{drwuEBMAL2016}, etc.

In this paper, we apply RG and tangent space features to supervised BCI regression problems. To overcome the limitation pointed out by Yger, Lotte, and Sugiyama \cite{Yger2015}, i.e., RG is less efficient when the dimensionality of the covariance matrix is large, we adopt an approach similar to what Barachant used in \cite{Barachant2014b}: we first use a spatial filter proposed in \cite{drwuSF2017} to reduce the dimensionality of the covariance matrices and also to increase the EEG signal quality, and then extract the RG features in the Riemannian tangent space. We validate the performance of the proposed approach in reaction time (RT) estimation from EEG signals measured in a large-scale sustained-attention PVT \cite{Drummond2005}, which collected 143 sessions of data from 17 subjects in a 5-month period. To our knowledge, this is the first time that RG has been used in BCI regression problems.

The remainder of this paper is organized as follows: Section~\ref{sect:Filter} describes the spatial filter we proposed earlier for supervised BCI regression problems. Section~\ref{sect:RG} introduces RG and the tangent space features for BCI regression problems. Section~\ref{sect:exp} describes the experimental setup, RT and EEG data preprocessing techniques, and the procedure to evaluate the performances of different feature extraction methods. Section~\ref{sect:results} presents the results of the comparative studies. Section~\ref{sect:discussions} provides parameter sensitivity analysis and additional discussions. Finally, Section~\ref{sect:conclusions} draws conclusions and outlines a future research direction.

\section{Spatial Filtering for Supervised BCI Regression Problems} \label{sect:Filter}

Recently we \cite{drwuSF2017} proposed two spatial filters for supervised BCI regression problems, which were extended from the common spatial pattern (CSP) algorithm for supervised classification problems. They have similar performance and computational cost. One of them, CSP for regression - one versus the rest (CSPR-OVR), is briefly introduced in this section, as the RG features are better extracted from the spatially filtered EEG data than the raw EEG data.

\begin{figure}[htpb]
\centering \includegraphics[width=8cm,clip]{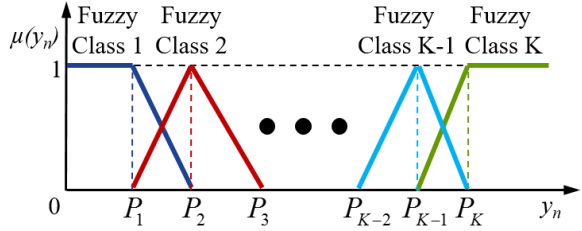} \caption{The $K$ fuzzy classes for $y_n$.} \label{fig:FSs}
\end{figure}

Let $\mathbf{X}_n\in \mathbb{R}^{C\times S}$ ($n=1,...,N$) denote the $n$th EEG trial in the training data, where $C$ is the number of channels and $S$ the number of time samples. We assume that the mean of each channel measurement has been removed, which is usually performed by band-pass filtering. Let $y_n\in\mathbb{R}$ be the corresponding RT of the $n$th trial. CSPR-OVR first constructs $K$ fuzzy sets \cite{Zadeh1965}, which partition the training samples into $K$ fuzzy classes. To do that, it partitions the interval $[0, 100]$ into $K+1$ equal intervals, and denotes the partition points as $\{p_k\}_{k=1,...,K}$. It is easy to obtain that
\begin{align}
p_k=\frac{100\cdot k}{K+1},\qquad k=1,...,K \label{eq:pk}
\end{align}
For each $p_k$, CSPR-OVR then finds the corresponding $p_k$ percentile value of all training $y_n$ and denotes it as $P_k$. Next we define $K$ fuzzy classes from them, as shown in Fig.~\ref{fig:FSs}.

Then, for each fuzzy class, CSPR-OVR computes its mean spatial covariance matrix as:
\begin{align}
\bar{\mathbf{\Sigma}}_k=\frac{\sum_{n=1}^N \mu_k(y_n)\mathbf{X}_n\mathbf{X}_n^T}{\sum_{n=1}^N \mu_k(y_n)}, \qquad k=1,...,K\label{eq:fP0}
\end{align}
where $\mu_k(y_n)$ is the membership degree of $y_n$ in Fuzzy Class $k$.

Next CSPR-OVR designs a spatial filtering matrix $\mathbf{W}_k^*\in \mathbb{R}^{C\times F}$, where $F$ is the number of individual vector filters, to maximize the variance difference between Fuzzy Class $k$ and the rest, i.e.,
\begin{align}
    \mathbf{W}_k^*
    =\arg\max\limits_{\mathbf{W}\in \mathbb{R}^{C\times F}}\frac{\mathrm{Tr}(\mathbf{W}^T\bar{\mathbf{\Sigma}}_k\mathbf{W})}
    {\mathrm{Tr}[\mathbf{W}^T(\sum_{i\neq k}\bar{\mathbf{\Sigma}}_i)\mathbf{W}]} \label{eq:W1}
    \end{align}
where $\mathrm{Tr}(\cdot)$ is the trace of a matrix. (\ref{eq:W1}) is a generalized Rayleigh quotient \cite{Golub1996}, and the solution $\mathbf{W}_k^*$ is the concatenation of the $F$ eigenvectors associated with the $F$ largest eigenvalues of the matrix $(\sum_{i\neq k}\bar{\mathbf{\Sigma}}_i)^{-1}\bar{\mathbf{\Sigma}}_k$.

The final spatial filtering matrix $\mathbf{W}^*\in \mathbb{R}^{C\times KF}$ is the concatenation of all $\mathbf{W}_k^*$, i.e.,
\begin{align}
\mathbf{W}^*=[\mathbf{W}_1^*, \quad \ldots, \quad\mathbf{W}_K^*] \label{eq:W}
\end{align}
and the spatially filtered trial for $\mathbf{X}_n$ is:
\begin{align}
\mathbf{X}_n'={\mathbf{W}^*}^T\mathbf{X}_n,\quad n=1,...,N. \label{eq:Xi}
\end{align}

In summary, the complete CSPR-OVR algorithm for supervised BCI regression problems is shown in Algorithm~\ref{alg:SF3}.

\begin{algorithm}[h] 
\KwIn{EEG training examples $(\mathbf{X}_n,y_n)$, where $\mathbf{X}_n\in \mathbb{R}^{C\times S}$, $n=1,...,N$\; \\
\hspace*{10mm} $K$, the number of fuzzy classes for $y_n$\; \\
\hspace*{10mm} $F$, the number of spatial filters for each \\
\hspace*{12mm} fuzzy class.}
\KwOut{Spatially filtered EEG trials $\mathbf{X}_n'\in \mathbb{R}^{KF\times S}$.}
Band-pass filter each $\mathbf{X}_n$ to remove the mean of each channel\;
Compute $\{p_k\}_{k=1,...,K}$ in (\ref{eq:pk})\;
Compute the corresponding percentile values $\{P_k\}_{k=1,...,K}$ for $y_n$\;
Construct the $K$ fuzzy classes as shown in Fig.~\ref{fig:FSs}\;
Compute $\bar{\mathbf{\Sigma}}_k$ by (\ref{eq:fP0})\;
Compute $\mathbf{W}_k^*$ by (\ref{eq:W1})\;
Construct $\mathbf{W}^*$ by (\ref{eq:W})\;
\textbf{Return} $\mathbf{X}_n'$ by (\ref{eq:Xi})
\caption{The CSPR-OVR spatial filter for supervised BCI regression problems \cite{drwuSF2017}.} \label{alg:SF3}
\end{algorithm}

\section{RG and the Tangent Space Features} \label{sect:RG}

This section introduces the basics of RG, and an approach to extract the Riemannian tangent space features.

\subsection{Riemannian Geometry}

The RG approach for BCI works on the covariance matrices of EEG trials, which are symmetric positive-definite and form a differentiable Riemannian manifold $\mathcal{M}$ \cite{Forstner1999} with dimensionality $R(R+1)/2$, where $R$ is the number of rows (columns) of the covariance matrices. As a result, we need to use Riemannian metrics, instead of the traditional Euclidean metrics, which are more appropriate for flat spaces of vectors. Particularly, we are interested in the distance measure between two covariance matrices, as many machine learning methods rely on such distances.

The \emph{Riemannian distance} $\delta(\bar{\mathbf{\Sigma}},\mathbf{\Sigma}_n)$ between two covariance matrices $\bar{\mathbf{\Sigma}}\in\mathbb{R}^{R\times R}$ and $\mathbf{\Sigma}_n\in\mathbb{R}^{R\times R}$, called the \emph{geodesic}, is the minimum length of a curve connecting them on the manifold $\mathcal{M}$. It can be computed as \cite{Moakher2005,Arsigny2007}:
\begin{align}
\delta(\bar{\mathbf{\Sigma}},\mathbf{\Sigma}_n)=\left\|\log \left(\bar{\mathbf{\Sigma}}^{-1}\mathbf{\Sigma}_n\right)\right\|_F
=\left[\sum_{r=1}^R\log^2\lambda_r\right]^{\frac{1}{2}}
\end{align}
where the subscript $_F$ denotes the Frobenius norm, and $\lambda_r$, $r=1,...,R$, are the real eigenvalues of $\bar{\mathbf{\Sigma}}^{-1}\mathbf{\Sigma}_n$.

At $\bar{\mathbf{\Sigma}}\in \mathcal{M}$, a scalar product can be defined in the associated \emph{tangent space} $\mathcal{T}_{\bar{\mathbf{\Sigma}}}\mathcal{M}$. This tangent space is Euclidean and locally homomorphic to the manifold. So, Riemannian distance computations in the manifold can be approximated by Euclidean distance computations in the tangent space \cite{Barachant2013}.

The \emph{logarithmic map} projects locally a $\mathbf{\Sigma}_n\in \mathcal{M}$ onto the tangent space $\mathcal{T}_{\bar{\mathbf{\Sigma}}}\mathcal{M}$ of $\bar{\mathbf{\Sigma}}$ by:
\begin{align}
\hat{\mathbf{\Sigma}}_n=\mathrm{Log}_{\bar{\mathbf{\Sigma}}}(\mathbf{\Sigma}_n)=
\bar{\mathbf{\Sigma}}^{\frac{1}{2}}\mathrm{logm}\left(\bar{\mathbf{\Sigma}}^{-\frac{1}{2}}
\mathbf{\Sigma}_n\bar{\mathbf{\Sigma}}^{-\frac{1}{2}}\right)\bar{\mathbf{\Sigma}}^{\frac{1}{2}} \label{eq:TS}
\end{align}
where $\mathrm{logm}(\cdot)$ denotes the logarithm of a matrix \cite{Berger2007}. The logarithm of a diagonalizable matrix $\mathbf{A=VDV}^{-1}$ is defined as $\mathrm{logm}(\mathbf{A})=\mathbf{VD}'\mathbf{V}^{-1}$, where $\mathbf{D}'$ is a diagonal matrix with elements $\mathbf{D}'_{i,i}=\log(\mathbf{D}_{i,i})$.

The \emph{exponential map} projects an element $\hat{\mathbf{\Sigma}}_n$ on the tangent space $\mathcal{T}_{\bar{\mathbf{\Sigma}}}\mathcal{M}$ back to the manifold $\mathcal{M}$ by:
\begin{align}
\mathbf{\Sigma}_n=\mathrm{Exp}_{\bar{\mathbf{\Sigma}}}(\hat{\mathbf{\Sigma}}_n)=
\bar{\mathbf{\Sigma}}^{\frac{1}{2}}\mathrm{expm}\left(\bar{\mathbf{\Sigma}}^{-\frac{1}{2}}
\mathbf{\Sigma}_n\bar{\mathbf{\Sigma}}^{-\frac{1}{2}}\right)
\bar{\mathbf{\Sigma}}^{\frac{1}{2}}
\end{align}
where $\mathrm{expm}(\cdot)$ denotes the exponential of a matrix \cite{Berger2007}. The exponential of a diagonalizable matrix $\mathbf{A=VDV}^{-1}$ is defined as $\mathrm{expm}(\mathbf{A})=\mathbf{VD}'\mathbf{V}^{-1}$, where $\mathbf{D}'$ is a diagonal matrix with elements $\mathbf{D}'_{i,i}=\exp(\mathbf{D}_{i,i})$.

\begin{figure}[htpb]\centering
\includegraphics[width=.9\linewidth,clip]{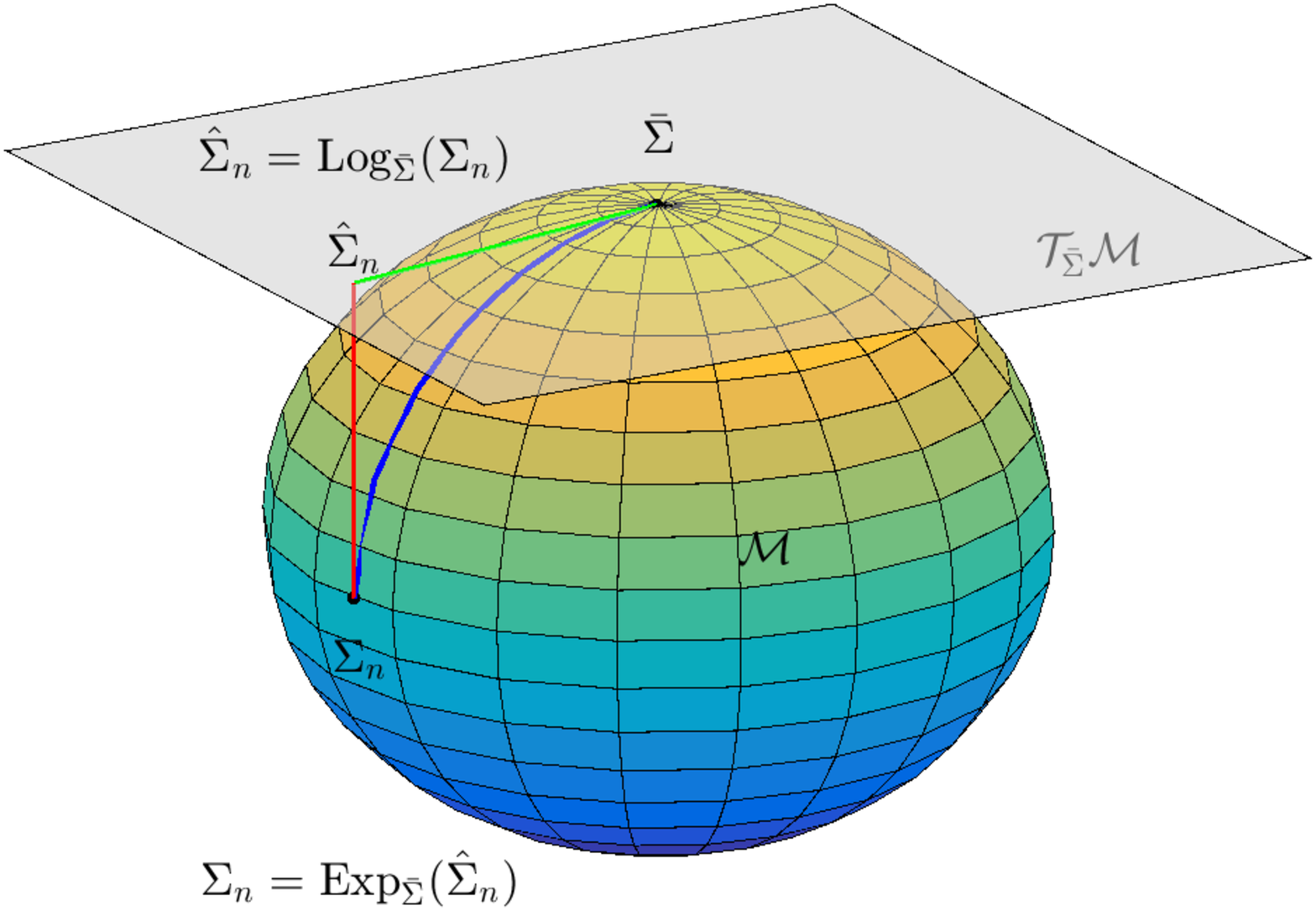}
\caption{Illustration of a manifold $\mathcal{M}$ and the corresponding local tangent space $\mathcal{T}_{\bar{\mathbf{\Sigma}}}\mathcal{M}$ at $\bar{\mathbf{\Sigma}}$. $\mathrm{Log}_{\bar{\mathbf{\Sigma}}}(\mathbf{\Sigma}_n)$ projects the matrix $\mathbf{\Sigma}_n$ on the manifold into the matrix $\hat{\mathbf{\Sigma}}_n$ in the tangent space of $\bar{\mathbf{\Sigma}}$. $\mathrm{Exp}_{\bar{\mathbf{\Sigma}}}(\hat{\mathbf{\Sigma}}_n)$ projects $\hat{\mathbf{\Sigma}}_n$ in the tangent space of $\bar{\mathbf{\Sigma}}$ into $\mathbf{\Sigma}_n$ on the manifold. The blue curve represents the geodesic between $\bar{\mathbf{\Sigma}}$ and $\mathbf{\Sigma}_n$ on the manifold.} \label{fig:RG}
\end{figure}

Fig.~\ref{fig:RG} illustrates a Riemannian manifold $\mathcal{M}$, the tangent space $\mathcal{T}_{\bar{\mathbf{\Sigma}}}\mathcal{M}$ at $\bar{\mathbf{\Sigma}}$, the geodesic between $\bar{\mathbf{\Sigma}}$ and $\mathbf{\Sigma}_n$, and the corresponding logarithmic and exponential maps.

The Riemannian distance $\delta(\bar{\mathbf{\Sigma}},\mathbf{\Sigma}_n)$ between two covariance matrices $\bar{\mathbf{\Sigma}}$ and $\mathbf{\Sigma}_n$ on the manifold $\mathcal{M}$ can also be computed by a Euclidean distance in the tangent space around $\bar{\mathbf{\Sigma}}$, i.e. \cite{Barachant2012},
\begin{align}
\delta(\bar{\mathbf{\Sigma}},\mathbf{\Sigma}_n)&=\left\|\mathrm{Log}_{\bar{\mathbf{\Sigma}}}
(\mathbf{\Sigma}_n)\right\|_{\bar{\mathbf{\Sigma}}}
=\left\|\mathrm{upper}\left(\bar{\mathbf{\Sigma}}^{-\frac{1}{2}}\hat{\mathbf{\Sigma}}_n
\bar{\mathbf{\Sigma}}^{-\frac{1}{2}}\right)\right\|_2\nonumber \\
&=\left\|\mathrm{upper}\left(\mathrm{logm}\left(\bar{\mathbf{\Sigma}}^{-\frac{1}{2}}
\mathbf{\Sigma}_n\bar{\mathbf{\Sigma}}^{-\frac{1}{2}}\right)\right)\right\|_2
\end{align}
where the $\mathrm{upper}(\cdot)$ operator keeps the upper triangular part of a symmetric matrix and vectorizes it by applying weight 1 for the diagonal elements and weight $\sqrt{2}$ for the out-of-diagonal elements \cite{Tuzel2008}.

The \emph{RG mean} \cite{Pennec2006a}, or the \emph{intrinsic mean} \cite{Fletcher2004}, of $N$ covariance matrices is defined as the matrix minimizing the sum of the squared Riemannian distances, i.e.,
\begin{align}
\bar{\mathbf{\Sigma}}\equiv \mathfrak{G}(\mathbf{\Sigma}_1,...,\mathbf{\Sigma}_N)
=\arg\min\limits_{\mathbf{\Sigma}}\sum_{n=1}^N\delta^2(\mathbf{\Sigma},\mathbf{\Sigma}_n)
\end{align}
There is no closed-form expression for the RG mean, but an iterative gradient descent algorithm (see Algorithm~\ref{alg:RGmean} \cite{Fletcher2004}) can be used to find the solution. Note that Algorithm~\ref{alg:RGmean} makes heavy use of the logarithmic and exponential maps. In this paper we used the implementation in the Matlab Covariance Toolbox\footnote{https://github.com/alexandrebarachant/covariancetoolbox.}.

\begin{algorithm}[h] 
\KwIn{$\mathbf{\Sigma}_n\in\mathbb{R}^{R\times R}$, $n=1,...,N$\;
\hspace*{10mm} $\epsilon >0$.}
\KwOut{The RG (intrinsic) mean $\bar{\mathbf{\Sigma}}\in\mathbb{R}^{R\times R}$.}
Initialize $\bar{\mathbf{\Sigma}}_0=\mathbf{0}\in\mathbb{R}^{R\times R}$, the zero matrix\;
Initialize $\bar{\mathbf{\Sigma}}=I\in\mathbb{R}^{R\times R}$, the identify matrix\;
\Repeat{$\left\|\bar{\mathbf{\Sigma}}-\bar{\mathbf{\Sigma}}_0\right\|<\epsilon$}{
$\bar{\mathbf{\Sigma}}_0=\bar{\mathbf{\Sigma}}$\;
$\hat{\mathbf{\Sigma}}=\frac{1}{N}\sum_{n=1}^N\mathrm{Log}_{\bar{\mathbf{\Sigma}}_0}(\mathbf{\Sigma}_n)$\;
$\bar{\mathbf{\Sigma}}=\mathrm{Exp}_{\bar{\mathbf{\Sigma}}_0}(\hat{\mathbf{\Sigma}})$.}
\textbf{Return} $\bar{\mathbf{\Sigma}}$
\caption{The gradient descent algorithm for computing the RG (intrinsic) mean \cite{Fletcher2004}.} \label{alg:RGmean}
\end{algorithm}

\subsection{Tangent Space Features for BCI Regression Problems}

To use the tangent space features for BCI regression problems, we first spatially filter each $\mathbf{X}_n$ to obtain $\mathbf{X}'_n$ in (\ref{eq:Xi}), and then estimate its spatial covariance matrix $\mathbf{\Sigma}_n\in\mathbb{R}^{KF\times KF}$ (note that each row of $\mathbf{X}'_n$ has zero mean):
\begin{align}
\mathbf{\Sigma}_n=\frac{1}{S}\mathbf{X}_n' \mathbf{X}_n'^T, \quad n=1,..., N \label{eq:Sigman}
\end{align}
Next, we compute the Riemannian mean $\bar{\mathbf{\Sigma}}$ of all $\mathbf{\Sigma}_n$ by Algorithm~\ref{alg:RGmean}, and take the $KF(KF+1)/2$ upper triangular part of $\mathrm{logm}\left(\bar{\mathbf{\Sigma}}^{-\frac{1}{2}}
\mathbf{\Sigma}_n\bar{\mathbf{\Sigma}}^{-\frac{1}{2}}\right)$ as our features. Note that we need to assign weight 1 to the diagonal elements of $\mathrm{logm}\left(\bar{\mathbf{\Sigma}}^{-\frac{1}{2}}
\mathbf{\Sigma}_n\bar{\mathbf{\Sigma}}^{-\frac{1}{2}}\right)$ and weight $\sqrt{2}$ to the out-of-diagonal elements so that their Euclidean norm is equal to the Riemannian distance between $\bar{\mathbf{\Sigma}}$ and $\mathbf{\Sigma}_k$. The weights do not have an effect when regression methods like LASSO are used, but are very important for distance based regression methods like kNN regression.

The complete tangent space feature extraction procedure for BCI regression problems is summarized in Algorithm~\ref{alg:RG}.

\begin{algorithm}[h] 
\KwIn{Spatially filtered EEG trial $\mathbf{X}_n'\in\mathbb{R}^{KF\times S}$, $n=1,...,N$.}
\KwOut{$KF(KF+1)/2$ tangent space features for each trial.}
Compute $\mathbf{\Sigma}_n$ by (\ref{eq:Sigman})\;
Compute $\bar{\mathbf{\Sigma}}$ by Algorithm~\ref{alg:RGmean}\;
Construct the $KF(KF+1)/2$ tangent space features for $\mathbf{X}_n'$ from $\mathrm{logm}\left(\bar{\mathbf{\Sigma}}^{-\frac{1}{2}}
\mathbf{\Sigma}_n\bar{\mathbf{\Sigma}}^{-\frac{1}{2}}\right)$.
\caption{The Riemannian tangent space feature extraction procedure for BCI regression problems.} \label{alg:RG}
\end{algorithm}

\section{Experiments and the Performance Evaluation Process} \label{sect:exp}

This section introduces a PVT experiment that was used to evaluate the performances of the proposed tangent space feature extraction method, and the corresponding RT and EEG data preprocessing procedures.

\subsection{Experiment Setup} \label{sect:PVT}

Seventeen university students (13 males; average age 22.4, standard deviation 1.6) from National Chiao Tung University (NCTU) in Taiwan volunteered to support the data-collection efforts over a 5-month period to study EEG correlates of attention and performance changes under specific conditions of real-world fatigue \cite{Kerick2016}, as determined by the percent effectiveness score of Readiband \cite{Russell2015}. The Institutional Review Board of NCTU approved the experimental protocol.

The customer-designed daily sampling system consists of a smartphone, actigraph, sleep diary, subjective scales of fatigue and stress, and software for recording, storing, transmitting, and analyzing data acquired from individuals in their natural environments on a daily basis. Each participant was provided a wrist-worn actigraph (Fatigue Science Readiband, Vancouver, BC), and was instructed to complete several subjective report scales and enter the percent effectiveness score from the actigraph approximately 30-60 minutes upon awakening each morning and to be available for experiment testing approximately once every 1-3 weeks over a 5-month period for a total of nine repeated sessions. Data recorded by the daily sampling system included electronically-adapted visual analog scales of fatigue and stress, the Karolinska Sleepiness Scale \cite{Akerstedt1990}, and the Pittsburgh Sleep Diary \cite{Monk1994}. The daily sampling data were automatically uploaded from the smartphone to a designated secure server at NCTU on a daily basis. In this way we could track and identify periods when the participants were currently exhibiting low, normal, or high levels of fatigue based on the percent effectiveness score values ($>$90\%, $70-90\%$, $<$70\%, respectively). The goal was to examine the participants during experiment sessions three times within each of the three fatigue levels. Most participants finished all nine sessions.

When the participants reported to the laboratory, we measured their fatigue level on site again right before the experiment to make sure it was close to the fatigue state reported via the smartphone. Upon completion of the related questionnaires and the informed consent form, subjects performed a PVT, a dynamic attention-shifting task, a lane-keeping task, and selected surveys preceding each condition. EEG data were recorded at 1000 Hz using a 64-channel NeuroScan Quik-Cap system (62 EEG channels and 1 electrocardiogram channel). The ground was between FPZ and FZ, and the reference channels were A1 and A2 at the mastoids.

In this paper we focus on the PVT \cite{Dinges1985}, which is a sustained-attention task that uses RT to measure the speed with which a subject responds to a visual stimulus. It is widely used, particularly by NASA, for its ease of scoring, simple metrics, convergent validity, and free of learning effects. In our experiment, the PVT was presented on a smartphone with each trial initiated as an empty solid white circle centered on the touchscreen that began to fill in red displayed as a clockwise sweeping motion like the hand of a clock. The sweeping motion was programmed to turn solid red in one second or terminate upon a response by the participants, which required them to tap the touchscreen with the thumb of their dominant hand. The RT was computed as the elapsed time between the appearance of the empty solid white circle and the participant's response. Following completion of each trial, the circle went back to solid white until the next trial. Inter-trial intervals consisted of random intervals between 2-10 seconds.

143 sessions of PVT data were collected from the 17 subjects, and each session lasted 10 minutes. Our goal is to predict the RT using a short EEG trial immediately before it.

\subsection{Performance Evaluation Process} \label{sect:process}

The following procedure was used to evaluate the performances of different feature extraction methods:
\begin{enumerate}
\item \emph{RT data preprocessing to remove outliers.}

The number of trials and the mean RTs for the 17 subjects are shown in Table~\ref{tab:RTs}. Subject 17 may have data recording issues, because many of his RTs were longer than 5 seconds, which are highly unlikely in practice, and his mean RT was more than two times larger than the largest mean RT from other subjects. So we excluded him from consideration in this paper, and only used Subjects 1-16.

\begin{table*}[ht] \centering \setlength{\tabcolsep}{1.6mm}
\caption{Number of trials and mean RTs for the 17 subjects.}   \label{tab:RTs}
\begin{tabular}{c|ccccccccccccccccc}   \hline
 Subject & 1 & 2 & 3 & 4 & 5 & 6 & 7 & 8 & 9 & 10 & 11 & 12 & 13 & 14 & 15 & 16   & 17\\ \hline
Number of trials & 809 & 813 & 839 & 685 & 843 & 465 & 833 & 610 & 769 & 813 & 743 & 828 & 803 &794 & 528 & 823 & 553 \\
Mean RT (s) & 0.47 &   0.51  &  0.47 &   0.98 &   0.49  &  0.42  &  0.61  &  0.44 &   0.81 &   0.46 &   1.03  &  0.45  &  0.57 &     0.59 &   0.67 &   0.49 &   2.14 \\ \hline
\end{tabular}
\end{table*}

The RTs were very noisy, and there were obvious outliers. It is very important to suppress the outliers and noise so that the performances of different algorithms can be more accurately compared. We employed the following 2-step procedure for RT data preprocessing:
\begin{enumerate}
\item \emph{Outlier removal}, which aimed to remove abnormally large RTs. First, a threshold $\theta=m_y+3\sigma_y$ was computed for each subject, where $m_y$ is the mean RT from all sessions of that subject, and $\sigma_y$ is the corresponding standard deviation. Then, all RTs larger than $\theta$ were removed. Note that the threshold was different for different subjects.
\item \emph{Moving average smoothing}, which replaced each RT by the average RT during a 60 seconds moving window centered at the onset of the corresponding PVT to suppress the noise.
\end{enumerate}

\item \emph{EEG data preprocessing to remove or suppress artifacts and noise.}

Generally raw EEG data recorded from the scalp contain many artifacts (e.g., head motion, blinks, eye movements, etc.) and noise (e.g., power-line noise, noise caused by changes in electrode impedances, etc.) \cite{Uriguen2015,Bigdely-Shamlo2015}, so it is very important to remove or suppress them to increase the signal-to-noise ratio before a machine learning algorithm is applied. This paper used the standardized early-stage EEG processing pipeline (PREP) \cite{Bigdely-Shamlo2015}, which consists of three steps: a) remove line-noise, b) determine and remove a robust reference signal, and, c) interpolate the bad channels (channels with a low recording signal-to-noise ratio).

The preprocessed EEG signals coming out of PREP were downsampled to 250 Hz. They were then epoched to 5-second trials according to the onset of the PVTs: if a PVT started at $t$, then the 62-channel EEG trial in $[t-5, t]$ seconds was used to predict the RT, i.e., $\mathbf{X}_n\in\mathbb{R}^{62\times 1250}$. Each trial was then individually filtered by a $[1, 20]$ Hz finite impulse response band-pass filter to make each channel zero-mean and to remove non-relevant high frequency components.

\item \emph{5-fold cross-validation to compute the regression performance for each combination of feature set and regression method.}

    We first randomly partitioned the trials into five folds; then, used four folds for supervised spatial filtering and regression model training, and the remaining fold for testing. We repeated this five times so that every fold was used in testing. Finally we computed the regression performances in terms of root mean square error (RMSE) and correlation coefficient (CC).

We extracted the following three different feature sets for each preprocessed EEG trial:
\begin{itemize}
\item \emph{Feature Set 1 (\texttt{FS1}): Theta and Alpha powerband features from the band-pass filtered EEG trials.} We computed the average power spectral density in the Theta band (4-8 Hz) and Alpha band (8-13 Hz) for each channel using Welch's method \cite{Welch1967}, and converted these $62\times 2=124$ band powers to dBs as our features.
\item \emph{Feature Set 2 (\texttt{FS2}): Theta and Alpha powerband features from EEG trials filtered by Algorithm~\ref{alg:SF3}.} This procedure was almost identical to the above one, except that the band-pass filtered EEG trials were also spatially filtered by Algorithm~\ref{alg:SF3} before the powerband features were computed. We used 3 fuzzy sets for the RTs, and 10 spatial filters for each fuzzy class, so that the spatially filtered EEG trials had dimensionality $30\times 1250$, and \texttt{FS2} had 60 dimensions.
\item \emph{Feature Set 3 (\texttt{FS3}): Riemannian tangent space features from EEG trials filtered by Algorithm~\ref{alg:SF3}.} That is, we first band-pass filtered the raw EEG signals, then spatially filtered them by Algorithm~1 ($K=10$ and $F=3$), and further applied Algorithm~\ref{alg:RG} to extract the tangent space features, which had $30\times 31/2=465$ dimensions.
\end{itemize}

Two regression methods were used on each feature set: LASSO \cite{Tibshirani1996}, and kNN regression \cite{Altman1992}.

For labeled training data $\{\mathbf{x}_n, y_n\}_{n=1,...,N}$, LASSO solves the following minimization problem to find a sparse linear regression model:
\begin{align}
\min_{\beta_0,\boldsymbol{\beta}}\left[\frac{1}{2N}\sum_{n=1}^N
\left(y_n-\beta_0-\boldsymbol{\beta}^T\mathbf{x}_n\right)^2+\lambda\left\|\boldsymbol{\beta}\right\|_1\right]
\end{align}
where $\lambda> 0$ is an adjustable parameter, which was optimized by an inner 5-fold cross-validation on the training dataset in this paper. Once $\beta_0$ and $\boldsymbol{\beta}$ are identified, the final LASSO regression model is:
\begin{align}
\hat{y}_n=\beta_0+\boldsymbol{\beta}^T\mathbf{x}_n
\end{align}

We used $k=5$ in kNN. Once the five nearest neighbors $\{\mathbf{x}_i, y_i\}_{i=1,...,5}$ to the new trial $\mathbf{x}_n$ are identified, the regression output is computed as a weighted average:
\begin{align}
\hat{y}_n=\frac{\sum_{i=1}^5 w_iy_i}{\sum_{i=1}^5 w_i}
\end{align}
where the weights are the inverses of the feature distances:
\begin{align}
w_i=\frac{1}{\|\mathbf{x}_n-\mathbf{x}_i\|_2}
\end{align}

\item \emph{Repeat Step~3 10 times and compute the average regression performance.}
\end{enumerate}

\section{Experimental Results} \label{sect:results}

This section compares the informativeness of the features in \texttt{FS1}, \texttt{FS2} and \texttt{FS3}, and presents the regression performances.

\subsection{Informativeness of the Features}

Before studying the regression performance, it is important to check if the extracted features in \texttt{FS1}, \texttt{FS2} and \texttt{FS3} are indeed meaningful.

\begin{figure}[htpb]\centering
\subfigure[]{\includegraphics[width=.49\linewidth,clip]{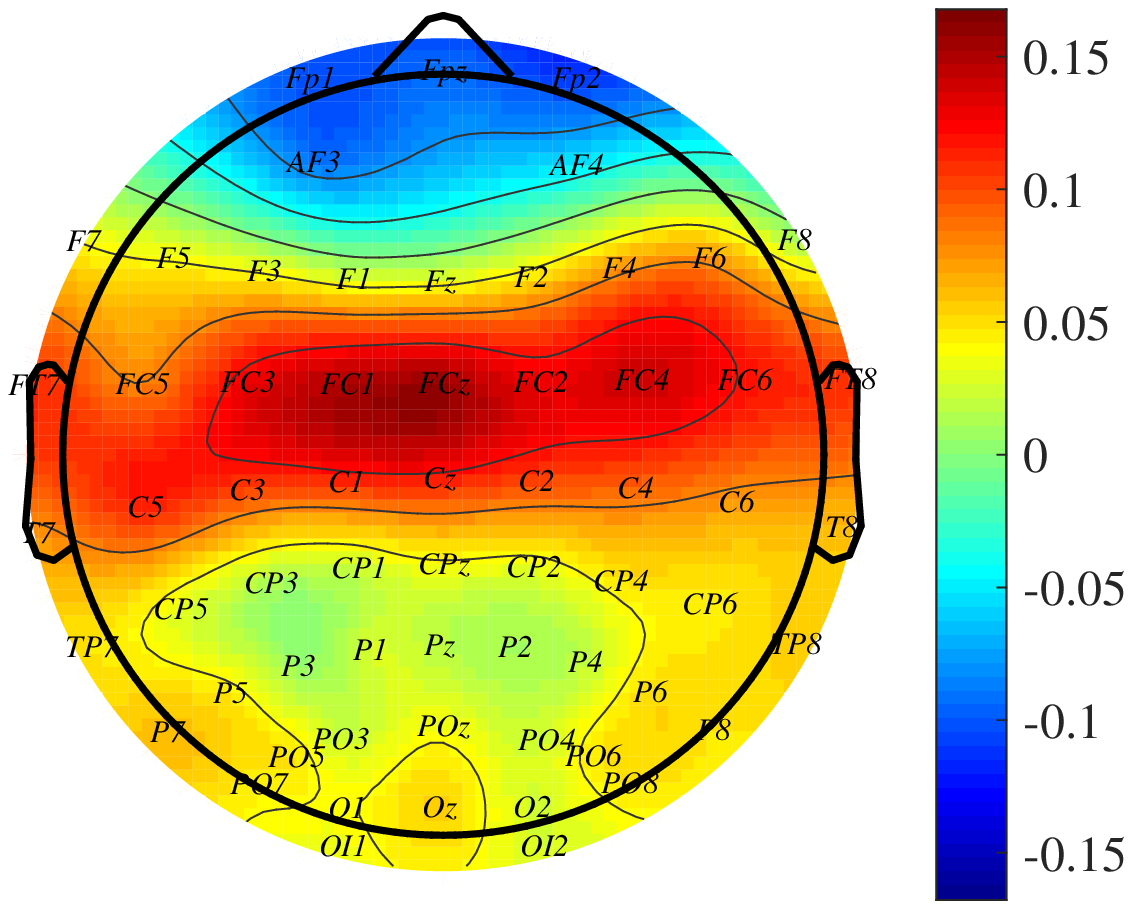}}
\subfigure[]{\includegraphics[width=.49\linewidth,clip]{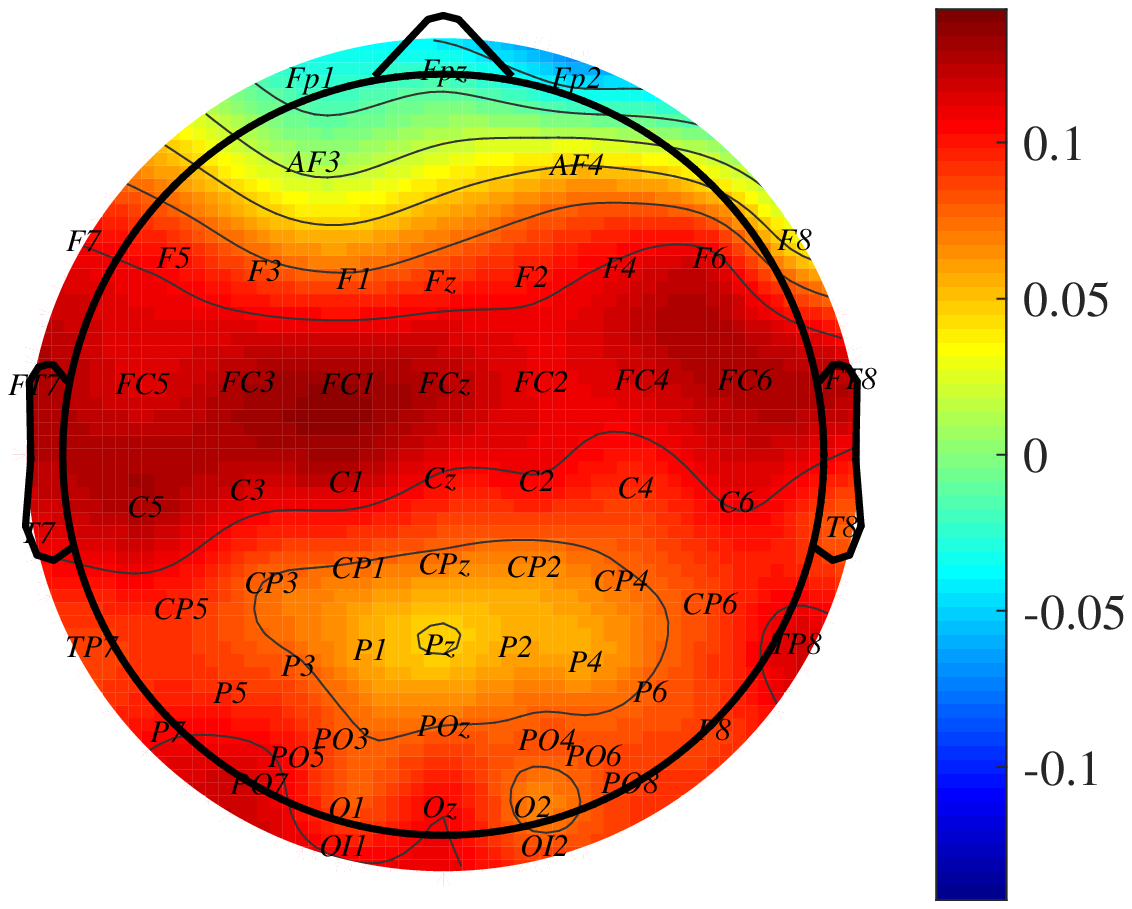}}
\caption{Topoplot of the average CC between the RT and the powerband features from \texttt{FS1} at different channel locations. (a) theta; (b) alpha.} \label{fig:topo}
\end{figure}

In this first study, we computed the CC between the RT and powerband features in \texttt{FS1} at different channel locations for each of the 16 subjects, and then averaged them. The corresponding topoplot is shown in Fig.~\ref{fig:topo}. Both theta and alpha band powers show higher correlation at the central and central-frontal regions of the brain; however, generally the CC is small. This indicates that \texttt{FS1} features are not very informative.

\begin{figure}[htpb]\centering
\includegraphics[width=\linewidth,clip]{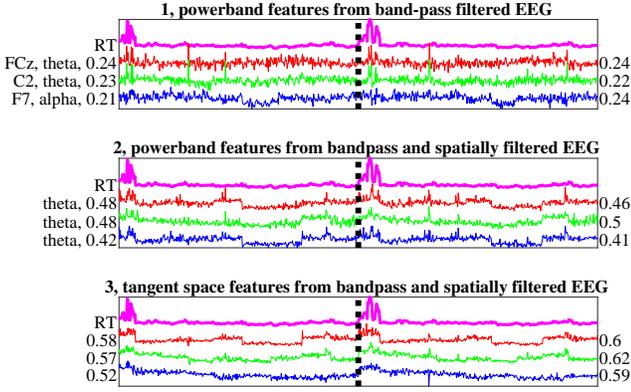}
\caption{Features from different feature extraction methods, and the corresponding training and testing CCs with the RT.} \label{fig:corr}
\end{figure}

In the second study, we picked a typical subject, partitioned his data randomly into 50\% training and 50\% testing, and extracted the powerband features \texttt{FS1}. We then designed the spatial filters using Algorithm~1 on the training data, and extracted the corresponding powerband features \texttt{FS2}, and the Riemannian tangent space features \texttt{FS3} using Algorithm~3. For each feature set, we identified the top three features that had the maximum CCs with the RT using the training data, and also computed the corresponding CCs for the testing data. The results are shown in Fig.~\ref{fig:corr}, where in each panel the data on the left of the black dotted line were used for training, and the right for testing. The top thick curve is the RT, and the bottom three curves are the maximally correlated features identified from the training data. The training and testing CCs are shown on the left and right of the corresponding feature, respectively. For \texttt{FS1}, we also show the corresponding channel labels and powerband names. For \texttt{FS2}, we only show the powerband names of the top three features, as a channel here does not have a specific label (each channel in \texttt{FS2} is a weighted combination of all 62 physical electrodes). Fig.~\ref{fig:corr} shows that \texttt{FS2} gave much smoother features than \texttt{FS1}, and also achieved much larger CCs to the RT, both in training and testing, suggesting that spatial filtering by Algorithm~1 can indeed increase the signal quality. \texttt{FS3} further achieved larger training and testing CCs to the RT than \texttt{FS2}, suggesting that the tangent space features are more informative than the powerband features.

\subsection{Estimation Performance Comparison}

\begin{figure}[htpb]\centering
\includegraphics[width=\linewidth,clip]{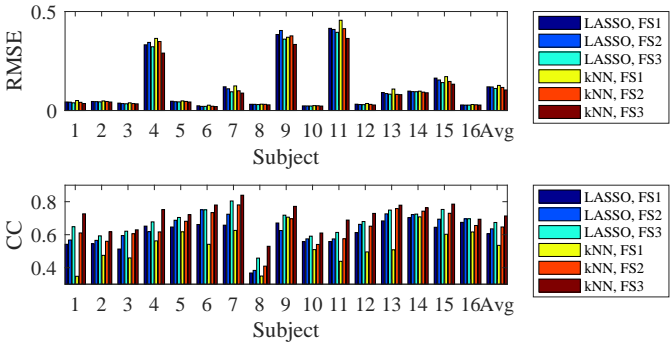}
\caption{RMSEs and CCs of the six approaches on the 16 subjects. } \label{fig:perf}
\end{figure}

The RMSEs and CCs of LASSO and kNN using three different feature sets are shown in Fig.~\ref{fig:perf} for the 16 subjects. Recall that for each subject the feature extraction methods were run 10 times, each with randomly partitioned training and testing data, and the average regression performances are shown here. The average RMSEs and CCs across all subjects are also shown in the last group of each panel.

Fig.~\ref{fig:perf} shows that regardless of which regression method was used, generally \texttt{FS2} resulted in smaller RMSEs and larger CCs than \texttt{FS1}, suggesting that the spatial filtering approach can indeed improve the regression performance. Fig.~\ref{fig:perf} also shows that \texttt{FS3} further achieved better RMSEs and CCs than \texttt{FS2}, suggesting that the tangent space features were more effective than the powerband features. Finally, LASSO had better performance than kNN on \texttt{FS1}, but kNN became better on \texttt{FS2} and \texttt{FS3}. The RMSEs for Subjects~4, 9 and 11 in Fig.~\ref{fig:perf} are much larger than others, because, as shown in Table~\ref{tab:RTs}, these three subjects have much larger RTs than others.

\begin{figure}[htpb]\centering
\includegraphics[width=\linewidth,clip]{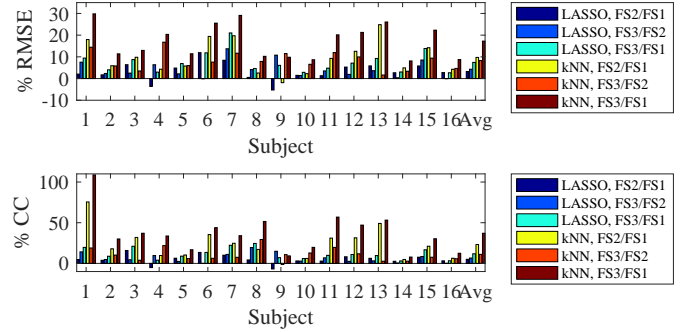}
\caption{Pairwise percentage performance improvement of the algorithms on the 16 subjects. } \label{fig:prc}
\end{figure}

To illustrate the performance differences among the three feature extraction methods from another viewpoint, Fig.~\ref{fig:prc} shows the corresponding percentage performance improvements of LASSO and kNN using the three feature sets, where the legend ``\texttt{LASSO,FS2}/\texttt{FS1}" means the percentage performance improvement of LASSO on \texttt{FS2} over LASSO on \texttt{FS1}, and other legends should be understood in a similar manner. For LASSO, on average \texttt{FS3} had $4.30\%$ smaller RMSE than \texttt{FS2}, and $6.59\%$ larger CC. For kNN, on average \texttt{FS3} had $8.30\%$ smaller RMSE than \texttt{FS2}, and $11.13\%$ larger CC. These results again demonstrated that the tangent space features are more effective than the traditional powerband features.

We also performed a two-way Analysis of Variance (ANOVA) for different regression algorithms to check if the raw RMSE and CC differences among the three feature sets (\texttt{FS1}, \texttt{FS2}, and \texttt{FS3}) were statistically significant, by setting the subjects as a random effect. The results are shown in Table~\ref{tab:ANOVA} as ``$p$ for raw values." Study results showed that there were statistically significant differences (at 5\% level) in raw CCs among different feature sets for both LASSO and kNN, but not for raw RMSEs.

However, because the RTs from different subjects had significantly different magnitudes, an ANOVA on the raw RMSEs and CCs may be unfair for those subjects with small RTs. So, we also performed a two-way ANOVA for different algorithms and feature sets on the ratios. For example, to compute the RMSE ratios for LASSO, we replaced all RMSEs for \texttt{FS1} by 1, the RMSEs for \texttt{FS2} by the ratios of the corresponding RMSEs from \texttt{FS2} over those from \texttt{FS1}, and the RMSEs for \texttt{FS3} by the ratios of the corresponding RMSEs from \texttt{FS3} over those from \texttt{FS1}. In this way the RMSEs were normalized, and hence different subjects were treated equally. The corresponding ANOVA test results are shown in Table~\ref{tab:ANOVA} as ``$p$ for ratios." Observe that there were statistically significant differences (at 5\% level) in both RMSE ratios and CC ratios among different feature sets for both LASSO and kNN.

\begin{table}[!ht] \centering \setlength{\tabcolsep}{2mm}
\caption{$p$-values of two-way ANOVA tests for $\{\texttt{FS1},\ \texttt{FS2}, \ \texttt{FS3}\}$.}   \label{tab:ANOVA}
\begin{tabular}{l|cc||cc}   \hline
&\multicolumn{2}{c||}{LASSO} & \multicolumn{2}{c}{kNN} \\ \hline
&RMSE & CC & RMSE & CC \\ \hline
   $p$ for raw values & .8183 & $\mathbf{.0000}$ &.2742&  $\mathbf{.0000}$\\
   $p$ for ratios & $\mathbf{.0000}$ & $\mathbf{.0000}$ &$\mathbf{.0000}$&  $\mathbf{.0000}$\\ \hline
\end{tabular}
\end{table}

Then, non-parametric multiple comparison tests based on Dunn's procedure \cite{Dunn1961,Dunn1964} were used to determine if the difference between any pair of algorithms was statistically significant, with a $p$-value correction using the False Discovery Rate method \cite{Benjamini1995}. The $p$-values for the raw values are shown in Table~\ref{tab:Dunn1}, and the $p$-values for the ratios are shown in Table~\ref{tab:Dunn2}, where the statistically significant ones are marked in bold. Table~\ref{tab:Dunn1} shows that the raw RMSE difference between \texttt{FS3} and \texttt{FS1} was statistically significant when kNN was used. Furthermore, the raw CC differences between all pairs of feature sets were statistically significant. Table~\ref{tab:Dunn2} shows that the ratio differences between all pairs of feature sets were statistically significant, for both LASSO and kNN.

\begin{table}[!ht] \centering \setlength{\tabcolsep}{1mm}
\caption{$p$-values of non-parametric multiple comparison on the raw values for $\{\texttt{FS1},\ \texttt{FS2}, \ \texttt{FS3}\}$.}   \label{tab:Dunn1}
\begin{tabular}{l|cc|cc|cc|cc}   \hline
&\multicolumn{4}{c|}{LASSO} & \multicolumn{4}{c}{kNN} \\ \hline
&\multicolumn{2}{c|}{RMSE} &\multicolumn{2}{c|}{CC} & \multicolumn{2}{c|}{RMSE} & \multicolumn{2}{c}{CC} \\ \hline
     &  \texttt{FS1} & \texttt{FS2} &  \texttt{FS1} & \texttt{FS2}  &  \texttt{FS1} & \texttt{FS2} &  \texttt{FS1} & \texttt{FS2}  \\ \hline
   \texttt{FS2} &  .2143 & & \textbf{.0001} &  &.0852 & & \textbf{.0000} &\\
  \texttt{FS3}  &.1319&.2702 &\textbf{.0000}& \textbf{.0001}&\textbf{.0034}&.0711 &\textbf{.0000}&\textbf{.0000}\\ \hline
\end{tabular}
\end{table}

\begin{table}[!ht] \centering \setlength{\tabcolsep}{1mm}
\caption{$p$-values of non-parametric multiple comparison on the ratios for $\{\texttt{FS1},\ \texttt{FS2}, \ \texttt{FS3}\}$.}   \label{tab:Dunn2}
\begin{tabular}{l|cc|cc|cc|cc}   \hline
&\multicolumn{4}{c|}{LASSO} & \multicolumn{4}{c}{kNN} \\ \hline
&\multicolumn{2}{c|}{RMSE} &\multicolumn{2}{c|}{CC} & \multicolumn{2}{c|}{RMSE} & \multicolumn{2}{c}{CC} \\ \hline
     &  \texttt{FS1} & \texttt{FS2} &  \texttt{FS1} & \texttt{FS2}  &  \texttt{FS1} & \texttt{FS2} &  \texttt{FS1} & \texttt{FS2}  \\ \hline
   \texttt{FS2} &  \textbf{.0000} & & \textbf{.0000} &  &\textbf{.0000} & & \textbf{.0000} &\\
  \texttt{FS3}  &\textbf{.0000}&\textbf{.0000} &\textbf{.0000}& \textbf{.0000}&\textbf{.0000}&\textbf{.0000} &\textbf{.0000}&\textbf{.0000}\\ \hline
\end{tabular}
\end{table}

\section{Discussions} \label{sect:discussions}

This section provides parameter sensitivity analysis and additional discussions.

\subsection{Parameter Sensitivity Analysis}

Tangent space feature extraction relies on the spatial filter in Algorithm~1, which has two adjustable parameters: $K$, the number of fuzzy classes for the RTs, and $F$, the number of spatial filters for each fuzzy class. The filtering performance is robust to $K$ but changes noticeably when $F$ changes \cite{drwuSF2017}. As a result, the performance of the tangent space features also varies as $F$ changes. In this subsection we study the sensitivity of the regression performance to $F$.

\begin{figure}[tb]\centering
\subfigure[]{\includegraphics[width=\linewidth,clip]{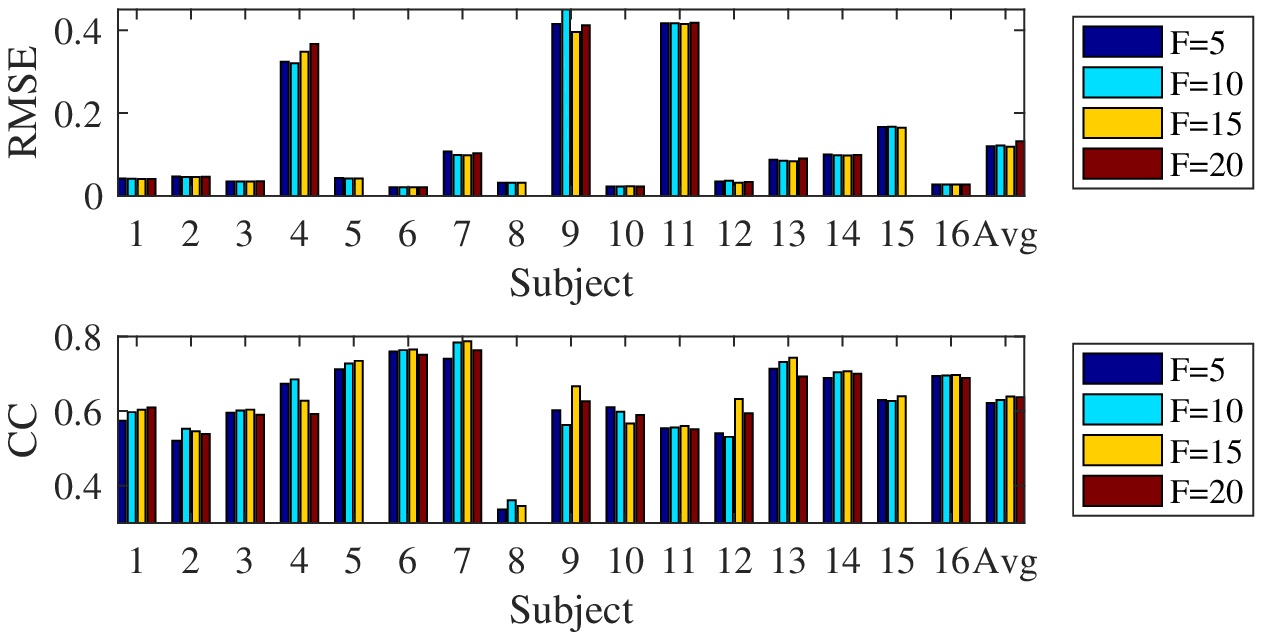}}
\subfigure[]{\includegraphics[width=\linewidth,clip]{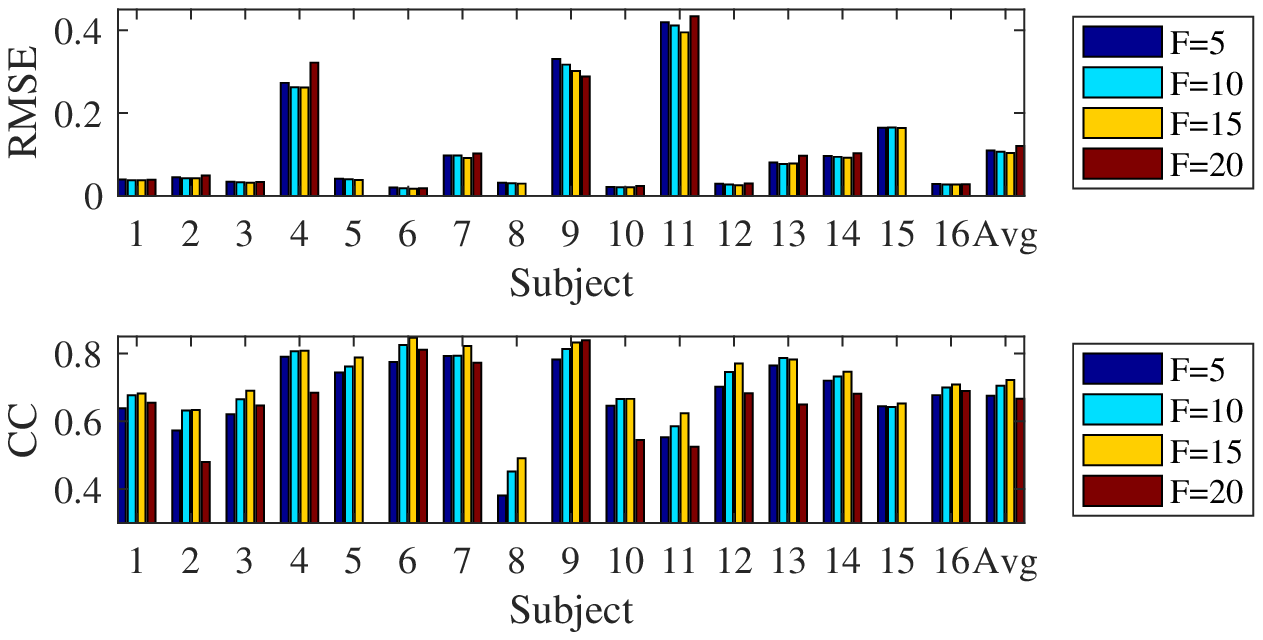}}
\caption{RMSEs and CCs of (a) LASSO and (b) kNN with respect to $F$, the number of spatial filters for each fuzzy class in Algorithm~1.} \label{fig:nFilters}
\end{figure}

The regression performances for $F=\{5, 10, 15, 20\}$ ($K$ was fixed to be 3) are shown in Fig.~\ref{fig:nFilters}. Algorithms~1 and 3 were repeated five times, each time with a random partition of training and testing data, and the average regression results are shown. Note that $F$ cannot be too large because of three constraints: 1) $F$ cannot exceed the number of channels ($C$) in the original EEG data, because $\bar{\mathbf{\Sigma}}_k\bar{\mathbf{\Sigma}}^{-1}\in \mathbb{R}^{C\times C}$ in (\ref{eq:W1}) has at most $C$ eigenvectors; 2) the tangent space features have dimensionality $KF(KF+1)/2$, which increases rapidly with $F$; so, a large $F$ can easily result in over-fitting; and, 3) there may be numerical difficulties in computing the RG mean when $F$ is large, e.g., for Subjects~5, 8 and 15 in Fig.~\ref{fig:nFilters} when $F=20$.

Fig.~\ref{fig:nFilters} shows that the regression performance increased when $F$ increased from 5 to 15, but decreased when $F$ further increased to 20. For the PVT experiment, $F\in[10,15]$ seemed to achieve a good compromise between performance and computational cost.

\begin{figure}[htpb]\centering
\subfigure[]{\includegraphics[width=\linewidth,clip]{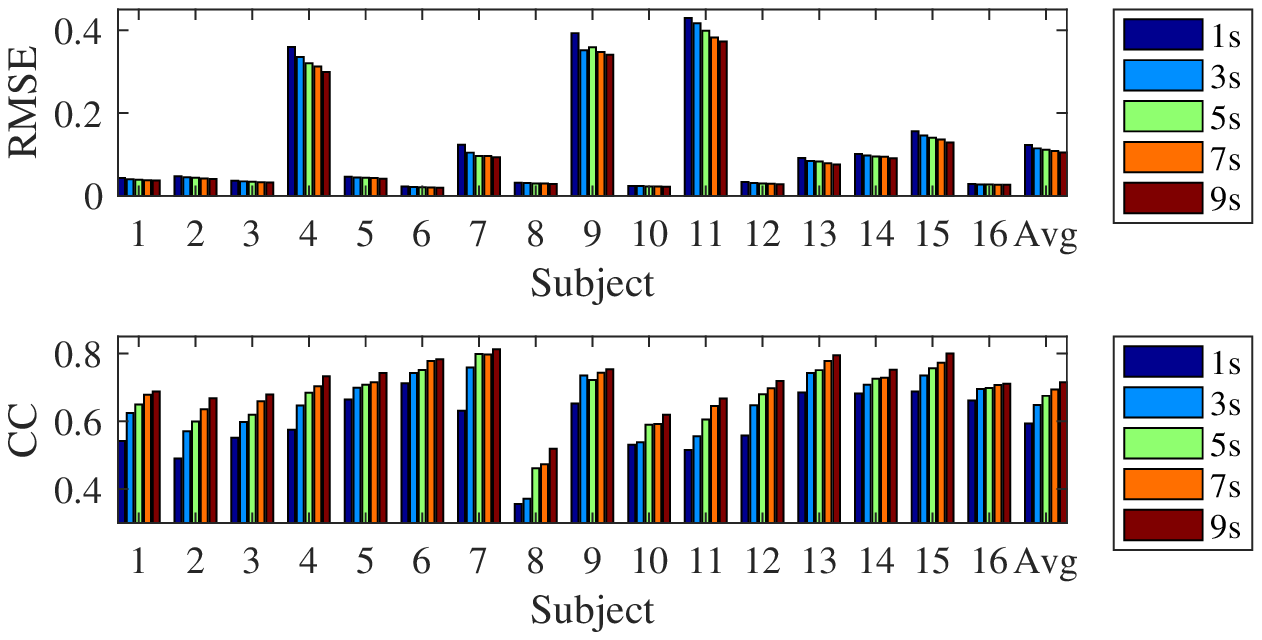}}
\subfigure[]{\includegraphics[width=\linewidth,clip]{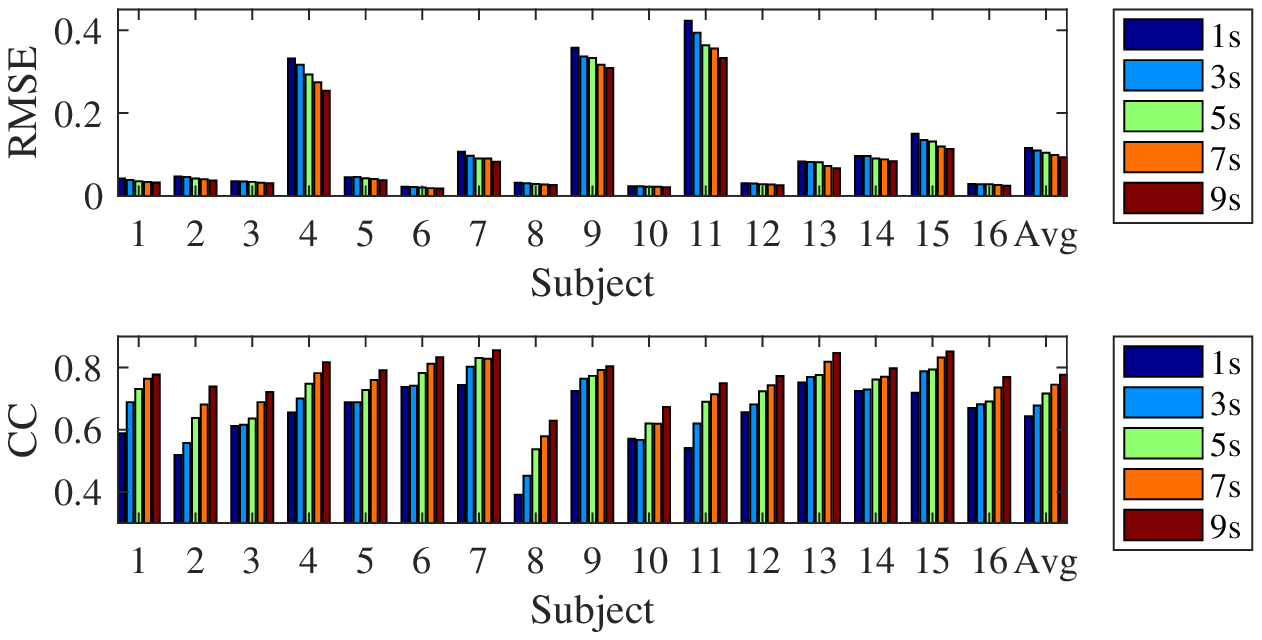}}
\caption{RMSEs and CCs of (a) LASSO and (b) kNN with respect to the trial length.} \label{fig:trialLength}
\end{figure}

Additionally, in the previous subsection we used 5-second EEG trials to estimate the corresponding RT, and it is also interesting to study how the estimation performance changes with different trial lengths. The results are shown in Fig.~\ref{fig:trialLength} for trial lengths of $\{1, 3, 5, 7, 9\}$ seconds. In general, as trial length increased, the estimation performance improved. However, a longer trial means heavier computational cost and larger delay in estimation. Furthermore, a trial cannot be arbitrary long, as then it cannot capture the up-to-date RT. These effects should be taken into consideration when choosing the right trial length.

\subsection{Regression Performance versus the Number of Features}

Recall from Section~\ref{sect:process} that \texttt{FS1} has 124 features, \texttt{FS2} has 60 features, and \texttt{FS3} has 465 features, i.e., \texttt{FS3} has much more features than \texttt{FS1} and \texttt{FS2}. So, \texttt{FS3}'s superior performance may be due to its increased number of features. In this subsection we investigate the relationship between the regression performance and the number of useful features.

\begin{figure}[htpb]\centering
\includegraphics[clip,width=\linewidth]{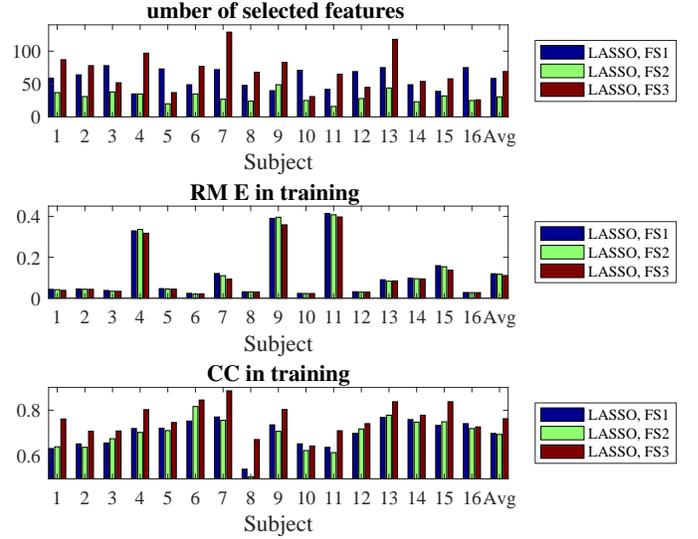} \caption{The nubmer of features selected by LASSO, and the corresponding training RMSEs and CCs.} \label{fig:numFeatures}
\end{figure}

Because LASSO automatically selects the most useful features, whereas kNN always uses all the features, in this study we focus only on LASSO. For each subject and each feature set, we used all data in LASSO training, and recorded the number of selected features, as well as the corresponding training RMSEs and CCs. The results are shown in Fig.~\ref{fig:numFeatures}. On average LASSO selected 58.6 features from \texttt{FS1}, 30.6 features from \texttt{FS2}, and 69.1 features from \texttt{FS3}. Although the selected \texttt{FS2} subset was only about half the size of the selected \texttt{FS1} subset, they resulted in similar overall training RMSEs and CCs. Connecting this observation with that in the previous subsection, i.e., \texttt{FS2} had much better testing RMSEs and CCs than \texttt{FS1}, we can conclude that the CSPR-OVR spatial filter can aggregate the most useful information into just a small number of features, which reduces overfitting and improves the generalization performance. Fig.~\ref{fig:numFeatures} also shows that the selected \texttt{FS3} subset was slightly larger than the selected \texttt{FS1} subset, but the \texttt{FS3} subset resulted in much better training performance, and also much better testing performance, as presented in the previous subsection. These observations together suggest that the Riemannian geometry approach can indeed extract some novel informative features, which improve both the training and the testing performances.

\subsection{Computational cost}

The training of our feature extraction method (\texttt{FS3}) consists of three steps: 1) design the CSPR-OVR filter by Algorithm~1; 2) compute the RG mean $\bar{\mathbf{\Sigma}}$ by Algorithm~2; and, 3) map the spatially filtered EEG trials to the Riemannian tangent space by Algorithm~3. Once the training is done, feature extraction for a testing trial can be performed very efficiently: a matrix multiplication (\ref{eq:Xi}) is first used to spatially filter it, and then another matrix multiplication (\ref{eq:Sigman}) is used to compute its spatial covariance matrix $\mathbf{\Sigma}_n$; finally, compute $\mathrm{logm}\left(\bar{\mathbf{\Sigma}}^{-\frac{1}{2}}
\mathbf{\Sigma}_n\bar{\mathbf{\Sigma}}^{-\frac{1}{2}}\right)$ and take its upper triangular part as the features. Note that $\bar{\mathbf{\Sigma}}$ has been obtained in training, so $\bar{\mathbf{\Sigma}}^{-\frac{1}{2}}$ can be pre-computed, and hence $\bar{\mathbf{\Sigma}}^{-\frac{1}{2}}
\mathbf{\Sigma}_n\bar{\mathbf{\Sigma}}^{-\frac{1}{2}}$ is also a simple matrix multiplication. So, in this subsection we focus on the training computational cost only.

\begin{figure}[htpb]\centering
\includegraphics[clip,width=.8\linewidth]{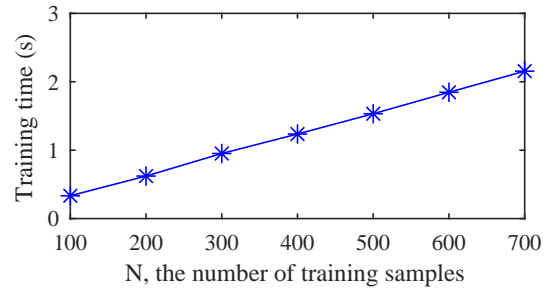} \caption{The training time of our feature extraction method w.r.t. $N$.} \label{fig:compCost}
\end{figure}

Let $N$ be the number of training samples. Then, the actual training time increased linearly with $N$, as shown in Fig.~\ref{fig:compCost}. The platform was a Dell XPS15 laptop (Intel i7-6700HQ CPU @2.60GHz, 16 GB memory) running Windows 10 Pro 64-bit and Matlab 2016b. A least squares curve fit shows that the training time is $0.0261+0.0030N$ seconds, which should not be a problem for a practical $N$.

\subsection{RT versus Fatigue State}

\begin{figure}[htpb]\centering
\includegraphics[clip,width=\linewidth]{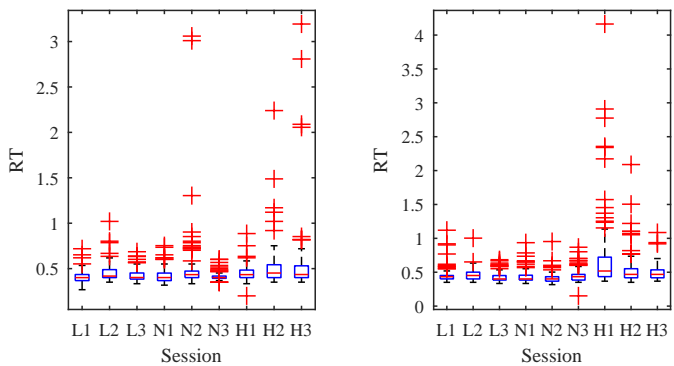} \caption{Boxplots of the RT in different fatigue states for two typical subjects.} \label{fig:RTstate}
\end{figure}

We also studied the relationship between the RT and the fatigue state. Our conjecture is that as the fatigue level goes up, the RT should be larger. Boxplots of the RT in different sessions for two typical subjects are shown in Fig.~\ref{fig:RTstate}, where ``L", ``N" and ``H" mean low, normal, and high fatigue, respectively. Fig.~\ref{fig:RTstate} shows that the mean RT of a high fatigue sessions is generally larger than that of a low or normal fatigue session, and the former also has more extreme values and a larger variance. The difference between a low fatigue session and a normal fatigue session is not obvious. These observations suggest that although the fatigue state contains some useful information, it may be too coarse for accurate RT prediction. That's why it was not used in this paper.

\section{Conclusions and Future Research} \label{sect:conclusions}

In this paper, we have proposed a new feature extraction approach for EEG-based BCI regression problems: a spatial filter is first used to increase the EEG trial signal quality and also to reduce the dimensionality of the covariance matrix, and then Riemannian tangent space features are extracted. We validated the performance of the proposed approach in RT estimation from EEG signals measured in a large-scale sustained-attention PVT experiment, and showed that compared with the traditional powerband features, the tangent space features can reduce the estimation RMSE by 4.30-8.30\%, and increase the estimation CC by 6.59-11.13\%. To our knowledge, this is the first time that RG has been used in BCI regression problems.

Our future research will focus on reducing the dimensionality of the tangent space features. As shown in Algorithm~3, the tangent space features have dimensionality $KF(KF+1)/2$, where $K$ is the number of fuzzy classes for the RTs, and $F$ is the number of spatial filters for each fuzzy class. So, the feature dimensionality increases quadratically with respect to both $K$ and $F$, which quickly results in overwhelming computational cost, overfitting, and numerical problems. We will investigate effective dimensionality reduction approaches for the tangent space features to reduce the computational cost while maintaining or even improving the regression performance.

\section*{Acknowledgement}

Research was sponsored by the U.S. Army Research Laboratory and was accomplished under Cooperative Agreement Numbers W911NF-10-2-0022 and W911NF-10-D-0002/TO 0023. The views and the conclusions contained in this document are those of the authors and should not be interpreted as representing the official policies, either expressed or implied, of the U.S. Army Research Laboratory or the U.S Government. This work was also partially supported by the Australian Research Council (ARC) under discovery grant DP150101645.


\begin{thebibliography}{10}

\bibitem{Akerstedt1990}
T.~Akerstedt and M.~Gillberg, ``Subjective and objective sleepiness in the
  active individual,'' \emph{International Journal of Neuroscience}, vol.~52,
  no. 1-2, pp. 29--37, 1990.

\bibitem{Altman1992}
N.~S. Altman, ``An introduction to kernel and nearest-neighbor nonparametric
  regression,'' \emph{The American Statistician}, vol.~46, no.~3, pp. 175--185,
  1992.

\bibitem{Amari2000}
S.~Amari, H.~Nagaoka, and D.~Harada, \emph{Methods of Information
  Geometry}.\hskip 1em plus 0.5em minus 0.4em\relax American Mathematical
  Society, 2000.

\bibitem{Arsigny2007}
V.~Arsigny, P.~Fillard, X.~Pennec, and N.~Ayache, ``Geometric means in a novel
  vector space structure on symmetric positive-definite matrices,''
  \emph{{SIAM} Journal on Matrix Analysis and Applications}, vol.~29, no.~1,
  pp. 328--347, 2007.

\bibitem{Barachant2012}
A.~Barachant, S.~Bonnet, M.~Congedo, and C.~Jutten, ``Multiclass brain-computer
  interface classification by {R}iemannian geometry,'' \emph{{IEEE} Trans. on
  Biomedical Engineering}, vol.~59, no.~4, pp. 920--928, 2012.

\bibitem{Barachant2013}
A.~Barachant, S.~Bonnet, M.~Congedo, and C.~Jutten, ``Classification of
  covariance matrices using a {R}iemannian-based kernel for {BCI}
  applications,'' \emph{Neurocomputing}, vol. 112, pp. 172--178, 2013.

\bibitem{Barachant2014b}
A.~Barachant. (2014) {MEG} decoding using {R}iemannian geometry and
  unsupervised classification. Accessed: 8/17/2016. [Online]. Available:
  \url{http://alexandre.barachant.org/wp-content/uploads/2014/08/documentation.pdf.}


\bibitem{Barachant2014}
A.~Barachant and M.~Congedo, ``A plug {\&} play {P300} {BCI} using information
  geometry,'' \emph{arXiv: 1409.0107}, 2014.

\bibitem{Benjamini1995}
Y.~Benjamini and Y.~Hochberg, ``Controlling the false discovery rate: A
  practical and powerful approach to multiple testing,'' \emph{Journal of the
  Royal Statistical Society, Series B (Methodological)}, vol.~57, pp. 289--300,
  1995.

\bibitem{Berger2007}
M.~Berger, \emph{A Panoramic View of {R}iemannian Geometry}.\hskip 1em plus
  0.5em minus 0.4em\relax New York, NY: Springer, 2007.

\bibitem{Bigdely-Shamlo2015}
N.~Bigdely-Shamlo, T.~Mullen, C.~Kothe, K.-M. Su, and K.~A. Robbins, ``The
  {PREP} pipeline: standardized preprocessing for large-scale {EEG} analysis,''
  \emph{Frontiers in Neuroinformatics}, vol.~9, 2015.

\bibitem{Bradberry2010}
T.~J. Bradberry, R.~J. Gentili, and J.~L. Contreras-Vidal, ``Reconstructing
  three-dimensional hand movements from noninvasive electroencephalographic
  signals,'' \emph{Journal of Neuroscience}, vol.~30, no.~9, pp. 3432--3437,
  2010.

\bibitem{Bradberry2011}
T.~J. Bradberry, R.~J. Gentili, and J.~L. Contreras-Vidal, ``Fast attainment of
  computer cursor control with noninvasively acquired brain signals,''
  \emph{Journal of Neural Engineering}, vol.~8, no.~3, 2011.

\bibitem{Congedo2013}
M.~Congedo, A.~Barachant, and A.~Andreev, ``A new generation of brain-computer
  interface based on {R}iemannian geometry,'' \emph{arXiv: 1310.8115}, 2013.

\bibitem{Dinges1985}
D.~F. Dinges and J.~W. Powell, ``Microcomputer analyses of performance on a
  portable, simple visual {RT} task during sustained operations,''
  \emph{Behavior research methods, instruments, \& computers}, vol.~17, no.~6,
  pp. 652--655, 1985.

\bibitem{Drummond2005}
S.~P. Drummond, A.~Bischoff-Grethe, D.~F. Dinges, L.~Ayalon, S.~C. Mednick, and
  M.~Meloy, ``The neural basis of the psychomotor vigilance task,''
  \emph{Sleep}, vol.~28, no.~9, pp. 1059--1068, 2005.

\bibitem{Dunn1961}
O.~Dunn, ``Multiple comparisons among means,'' \emph{Journal of the American
  Statistical Association}, vol.~56, pp. 62--64, 1961.

\bibitem{Dunn1964}
O.~Dunn, ``Multiple comparisons using rank sums,'' \emph{Technometrics},
  vol.~6, pp. 214--252, 1964.

\bibitem{Fletcher2004}
P.~T. Fletcher and S.~Joshi, ``Principal geodesic analysis on symmetric spaces:
  Statistics of diffusion tensors,'' \emph{Computer Vision and Mathematical
  Methods in Medical and Biomedical Image Analysis}, pp. 87--98, 2004.

\bibitem{Forstner1999}
W.~Forstner and B.~Moonen, ``A metric for covariance matrices,'' Dept. of
  Geodesy and Geoinformatics, Stuttgart University, Tech. Rep., 1999.

\bibitem{Fruitet2010}
J.~Fruitet, D.~J. McFarland, and J.~R. Wolpaw, ``A comparison of regression
  techniques for a two-dimensional sensorimotor rhythm-based brain-computer
  interface,'' \emph{Journal of Neural Engineering}, vol.~7, no.~1, 2010.

\bibitem{Golub1996}
G.~H. Golub and C.~F.~V. Loan, \emph{Matrix Computation}, 3rd~ed.\hskip 1em
  plus 0.5em minus 0.4em\relax Baltimore, MD: The Johns Hopkins University
  Press, 1996.

\bibitem{He2015}
B.~He, B.~Baxter, B.~J. Edelman, C.~C. Cline, and W.~W. Ye, ``Noninvasive
  brain-computer interfaces based on sensorimotor rhythms,'' \emph{Proc. of the
  {IEEE}}, vol. 103, no.~6, pp. 907--925, 2015.

\bibitem{Jayaram2016}
V.~Jayaram, M.~Alamgir, Y.~Altun, B.~Scholkopf, and M.~Grosse-Wentrup,
  ``Transfer learning in brain-computer interfaces,'' \emph{{IEEE}
  Computational Intelligence Magazine}, vol.~11, no.~1, pp. 20--31, 2016.

\bibitem{Kalunga2016}
E.~K. Kalunga, S.~Chevallier, K.~Djouani, E.~Monacelli, and Y.~Hamam, ``Online
  {SSVEP}-based {BCI} using {R}iemannian geometry,'' \emph{Neurocomputing},
  vol. 191, pp. 55--68, 2016.

\bibitem{Kerick2016}
S.~Kerick, C.-H. Chuang, J.-T. King, T.-P. Jung, J.~Brooks, B.~T. Files,
  K.~McDowell, and C.-T. Lin, ``Inter- and intra-individual variations in
  sleep, subjective fatigue, and vigilance task performance of students in
  their real-world environments over extended periods,'' 2016, submitted.

\bibitem{Lee2002}
J.~Lee, \emph{Introduction to Smooth Manifolds}.\hskip 1em plus 0.5em minus
  0.4em\relax New York, NY: Springer, 2002.

\bibitem{Li2012a}
Y.~Li, K.~Wong, and H.~de~Bruin, ``Electroencephalogram signals classification
  for sleep state decision -- a {R}iemannian geometry approach,'' \emph{{IET}
  Signal Processing}, vol.~6, no.~4, pp. 288--299, 2012.

\bibitem{Liao2012}
L.-D. Liao, C.-T. Lin, K.~McDowell, A.~Wickenden, K.~Gramann, T.-P. Jung, L.-W.
  Ko, and J.-Y. Chang, ``Biosensor technologies for augmented brain-computer
  interfaces in the next decades,'' \emph{Proc. of the {IEEE}}, vol. 100,
  no.~2, pp. 1553--1566, 2012.

\bibitem{Lin2005d}
C.~T. Lin, R.~C. Wu, S.~F. Liang, T.~Y. Huang, W.~H. Chao, Y.~J. Chen, and
  T.~P. Jung, ``{EEG}-based drowsiness estimation for safety driving using
  independent component analysis,'' \emph{{IEEE} Trans. on Circuits and
  Systems}, vol.~52, pp. 2726--2738, 2005.

\bibitem{Lin2008}
C.-T. Lin, Y.-C. Chen, T.-Y. Huang, T.-T. Chiu, L.-W. Ko, S.-F. Liang, H.-Y.
  Hsieh, S.-H. Hsu, and J.-R. Duann, ``Development of wireless brain computer
  interface with embedded multitask scheduling and its application on real-time
  driver's drowsiness detection and warning,'' \emph{{IEEE} Trans. on
  Biomedical Engineering}, vol.~55, no.~5, pp. 1582--1591, 2008.

\bibitem{Lin2006}
C.-T. Lin, L.-W. Ko, I.-F. Chung, T.-Y. Huang, Y.-C. Chen, T.-P. Jung, and
  S.-F. Liang, ``Adaptive {EEG}-based alertness estimation system by using
  {ICA}-based fuzzy neural networks,'' \emph{{IEEE} Trans. on Circuits and
  Systems-I}, vol.~53, no.~11, pp. 2469--2476, 2006.

\bibitem{Lotte2015}
F.~Lotte, ``Signal processing approaches to minimize or suppress calibration
  time in oscillatory activity-based brain-computer interfaces,'' \emph{Proc.
  of the {IEEE}}, vol. 103, no.~6, pp. 871--890, 2015.

\bibitem{Makeig2012}
S.~Makeig, C.~Kothe, T.~Mullen, N.~Bigdely-Shamlo, Z.~Zhang, and
  K.~Kreutz-Delgado, ``Evolving signal processing for brain-computer
  interfaces,'' \emph{Proc. of the {IEEE}}, vol. 100, no. Special Centennial
  Issue, pp. 1567--1584, 2012.

\bibitem{McFarland1997a}
D.~J. McFarland, A.~T. Lefkowicz, and J.~R. Wolpaw, ``Design and operation of
  an {EEG}-based brain-computer interface with digital signal processing
  technology,'' \emph{Behavior Research Methods, Instruments, \& Computers},
  vol.~29, no.~3, pp. 337--345, 1997.

\bibitem{McFarland2010}
D.~J. McFarland, W.~A. Sarnacki, and J.~R. Wolpaw, ``Electroencephalographic
  ({EEG}) control of three-dimensional movement,'' \emph{Journal of Neural
  Engineering}, vol.~7, no.~3, 2010.

\bibitem{Moakher2005}
M.~Moakher, ``A differential geometric approach to the geometric mean of
  symmetric positive-definite matrices,'' \emph{{SIAM} Journal on Matrix
  Analysis and Applications}, vol.~26, no.~3, pp. 735--747, 2005.

\bibitem{Monk1994}
T.~Monk, C.~Reynolds, D.~Kupfer, D.~Buysse, P.~Coble, A.~Hayes, M.~Machen,
  S.~Petrie, and A.~Ritenour, ``The {Pittsburgh} sleep diary,'' \emph{Journal
  of Sleep Research}, vol.~3, pp. 111--120, 1994.

\bibitem{Navarro-Sune2016}
X.~Navarro-Sune, A.~L. Hudson, F.~D.~V. Fallani, J.~Martinerie, A.~Witon,
  P.~Puget, M.~Raux, T.~Similowski, and M.~Chavez, ``Riemannian geometry
  applied to detection of respiratory states from {EEG} signals: the basis for
  a brain-ventilator interface,'' \emph{{IEEE} Trans. on Biomedical
  Engineering}, 2016, in press.

\bibitem{Parisi2014}
F.~Parisi, F.~Strino, B.~Nadler, and Y.~Kluger, ``Ranking and combining
  multiple predictors without labeled data,'' \emph{Proc. National Academy of
  Science ({PNAS})}, vol. 111, no.~4, pp. 1253--1258, 2014.

\bibitem{Pennec2006}
X.~Pennec, ``Intrinsic statistics on {R}iemannian manifolds: Basic tools for
  geometric measurements,'' \emph{Journal of Mathematical Imaging and Vision},
  vol.~25, no.~1, pp. 127--154, 2006.

\bibitem{Pennec2006a}
X.~Pennec, P.~Fillard, and N.~Ayache, ``A {R}iemannian framework for tensor
  computing,'' \emph{Int'l Journal of Computer Vision}, vol.~66, no.~1, pp.
  41--66, 2006.

\bibitem{Russell2015}
C.~Russell, J.~Caldwell, D.~Arand, L.~Myers, P.~Wubbels, and H.~Downs. (2015)
  Validation of the fatigue science readiband actigraph and associated
  sleep/wake classification algorithms. Accessed: 08/11/2016. [Online].
  Available:
  \url{http://static1.squarespace.com/static/550af02ae4b0cf85628d981a/t/5526c99ee4b019412c323758/14286053423
  03/Readiband_Validation.pdf.}

\bibitem{Tibshirani1996}
R.~Tibshirani, ``Regression shrinkage and selection via the lasso,''
  \emph{Journal of the Royal Statistical Society}, vol.~58, no.~1, pp.
  267--288, 1996.

\bibitem{Tuzel2008}
O.~Tuzel, F.~Porikli, and P.~Meer, ``Pedestrian detection via classification on
  {R}iemannian manifolds,'' \emph{{IEEE} Trans. on Pattern Analysis and Machine
  Intelligence}, vol.~30, no.~10, pp. 1713--1727, 2008.

\bibitem{Uriguen2015}
J.~A. Uriguen and B.~Garcia-Zapirain, ``{EEG} artifact removal --
  state-of-the-art and guidelines,'' \emph{Journal of Neural Engineering},
  vol.~12, no.~3, 2015.

\bibitem{Wang2015}
P.~Wang, J.~Lu, B.~Zhang, and Z.~Tang, ``A review on transfer learning for
  brain-computer interface classification,'' in \emph{Prof. 5th Int'l Conf. on
  Information Science and Technology (IC1ST)}, Changsha, China, April 2015.

\bibitem{Waytowich2016}
N.~R. Waytowich, V.~J. Lawhern, A.~W. Bohannon, K.~R. Ball, and B.~J. Lance,
  ``Spectral transfer learning using {I}nformation {G}eometry for a
  user-independent brain-computer interface,'' \emph{Frontiers in
  Neuroscience}, vol.~10, p. 430, 2016.

\bibitem{Wei2015}
C.-S. Wei, Y.-P. Lin, Y.-T. Wang, T.-P. Jung, N.~Bigdely-Shamlo, and C.-T. Lin,
  ``Selective transfer learning for {EEG}-based drowsiness detection,'' in
  \emph{Proc. {IEEE} Int'l Conf. on Systems, Man and Cybernetics}, Hong Kong,
  October 2015, pp. 3229--3232.

\bibitem{Welch1967}
P.~Welch, ``The use of fast {F}ourier transform for the estimation of power
  spectra: A method based on time averaging over short, modified
  periodograms,'' \emph{{IEEE} Trans. on Audio Electroacoustics}, vol.~15, pp.
  70--73, 1967.

\bibitem{Wolpaw1991}
J.~R. Wolpaw, D.~J. McFarland, G.~W. Neat, and C.~A. Forneris, ``An {EEG}-based
  brain-computer interface for cursor control,'' \emph{Electroencephalography
  and Clinical Neurophysiology}, vol.~78, no.~3, pp. 252--259, 1991.

\bibitem{Wolpaw2000}
J.~R. Wolpaw, D.~J. McFarland, and T.~M. Vaughan, ``Brain-computer interface
  research at the {W}adsworth {C}enter,'' \emph{{IEEE} Trans. on Rehabilitation
  Engineering}, vol.~8, no.~2, pp. 222--226, 2000.

\bibitem{Wolpaw2006}
J.~Wolpaw and N.~Birbaumer, ``Brain-computer interfaces for communication and
  control,'' in \emph{Textbook for Neural Repair and Repair and
  Rehabilitation}, M.~Selzer, L.~Cohen, F.~Gage, S.~Clarke, and P.~Duncan,
  Eds.\hskip 1em plus 0.5em minus 0.4em\relax Cambridge University Press, 2006,
  pp. 602--614.

\bibitem{drwuaBCI2015}
D.~Wu, C.-H. Chuang, and C.-T. Lin, ``Online driver's drowsiness estimation
  using domain adaptation with model fusion,'' in \emph{Proc. Int'l Conf. on
  Affective Computing and Intelligent Interaction}, Xi'an, China, September
  2015, pp. 904--910.

\bibitem{drwuSF2017}
D.~Wu, J.-T. King, C.-H. Chuang, C.-T. Lin, and T.-P. Jung, ``Spatial filtering
  for {EEG}-based regression problems in brain-computer interface ({BCI}),''
  \emph{{IEEE} Trans. on Fuzzy Systems}, 2017, accepted. [Online]. Available:
  \url{https://arxiv.org/abs/1702.02914}


\bibitem{drwuTFS2016}
D.~Wu, V.~J. Lawhern, S.~Gordon, B.~J. Lance, and C.-T. Lin, ``Driver
  drowsiness estimation from {EEG} signals using online weighted adaptation
  regularization for regression ({OwARR}),'' \emph{{IEEE} Trans. on Fuzzy
  Systems}, 2016, in press.

\bibitem{drwuEBMAL2016}
D.~Wu, V.~J. Lawhern, S.~Gordon, B.~J. Lance, and C.-T. Lin, ``Offline
  {EEG}-based driver drowsiness estimation using enhanced batch-mode active
  learning ({EBMAL}) for regression,'' in \emph{Proc. {IEEE} Int'l Conf. on
  Systems, Man and Cybernetics}, Budapest, Hungary, October 2016, pp. 730--736.

\bibitem{drwuSMLR2016}
D.~Wu, V.~J. Lawhern, S.~Gordon, B.~J. Lance, and C.-T. Lin, ``Spectral
  meta-learner for regression {(SMLR)} model aggregation: Towards
  calibrationless brain-computer interface ({BCI}),'' in \emph{Proc. {IEEE}
  Int'l Conf. on Systems, Man and Cybernetics}, Budapest, Hungary, October
  2016, pp. 743--749.

\bibitem{Yger2015}
F.~Yger, F.~Lotte, and M.~Sugiyama, ``Averaging covariance matrices for {EEG}
  signal classification based on the {CSP}: An empirical study,'' in
  \emph{Proc. 23rd European Signal Processing Conference (EUSIPCO)}, Nice,
  France, August 2015, pp. 2721--2725.

\bibitem{Zadeh1965}
L.~A. Zadeh, ``Fuzzy sets,'' \emph{Information and Control}, vol.~8, pp.
  338--353, 1965.

\end{thebibliography}
\end{document}